# Quantum computing for transport optimization

Christopher D B Bentley[1*], Samuel Marsh[1], André R R Carvalho[1], Philip Kilby[2],
Michael J Biercuk[1]

1. Q-CTRL, Sydney, NSW, Australia (christopher.bentley@q-ctrl.com)
2. CSIRO Data61, Canberra, ACT, Australia

**Abstract**

We explore the near-term intersection of quantum computing with the transport sector. To support near-term integration, we introduce a framework for assessing the suitability of transport optimization problems for obtaining potential performance enhancement using quantum algorithms. Given a suitable problem, we then present a workflow for obtaining valuable transport solutions using quantum computers, articulate the limitations on contemporary systems, and describe newly available performance-enhancing tools applicable to current commercial quantum computing systems. We make this integration process concrete by following the assessment framework and integration workflow for an exemplary vehicle routing optimization problem: the Capacitated Vehicle Routing Problem. We present novel advances to exponentially reduce the required computational resources, and experimentally demonstrate a prototype implementation exhibiting over 20X circuit performance enhancement on a real quantum device.

**Introducing quantum computing for transport solutions**

Transport operations rely on a variety of classical technologies where performance impacts the safety, efficiency and experience of passengers and customers. Advances in quantum technologies have demonstrated potential performance enhancements over classical counterparts in precision navigation and timing through quantum sensing, in battery design through quantum simulation, in communication security through quantum cryptography, and in transport optimization through quantum computing. While each of these applications are critical for transport applications, we focus on transport optimization, which underpins the planning and operation of all transport services [1].

A broad range of transport optimization problems, such as routing, scheduling, assignment and rostering, are *combinatorial optimization* problems that are hard to solve for classical computers (NP-hard optimization problems) [2]. The large scale or complexity of these problems - a common scenario for urban transport and logistics given the size of the networks and number of passengers or goods involved - results in the best classical algorithms either requiring impractical run-times, or falling short of the optimal solution. These limitations provide the appeal for tackling combinatorial optimization problems using quantum computing, which may be able to provide computational advantages for problems of this form.

Quantum computing operates via a different physical paradigm from classical computing; information is stored in quantum bits ('qubits'), and processing exploits physical effects such as quantum superposition and entanglement to deliver computational advantages in certain, limited cases. Indeed, algorithms such as Shor's factorization algorithm [3] and Grover's search algorithm [4] provide proven computational speed-ups over known classical algorithms.

In the context of optimization, Grover's search algorithm can be applied to find the global minimum of any discrete black-box function, and is proven to operate quadratically faster than is classically possible [4, 5]. Thus quantum computation can achieve provable theoretical (quadratic) advantage for brute-forcing hard combinatorial optimization problems that are relevant to transport. However the threshold of quantum advantage, when such a system will outperform the best classical alternative, generally mandates large systems beyond the reach of current engineering.

Small quantum hardware systems are now routinely available, and we have seen the first demonstrations that for a narrow range of mathematically interesting problems (that are otherwise detached from end-use applications) quantum computers may surpass the capabilities of their classical counterparts [6]. The current embryonic era of quantum hardware development has been labeled the Noisy Intermediate Scale Quantum (NISQ) era [7]; this description encapsulates limits on both hardware size (qubit count) and performance (error-prone devices), which in combination severely restrict the capabilities of today's machines, as well as the value of solutions that can be obtained using quantum algorithms.

In recognition of these challenges, a new class of algorithms has been developed with the objective of delivering potential computational advantage for practically relevant problems on heavily constrained hardware [8]. Quantum variational algorithms combine a small quantum coprocessor with classical computing infrastructure in order to dedicate valuable quantum hardware to the most challenging aspects of a calculation. In cases where problems involve the optimization of quantum or classical systems, including certain transport problems, it may be possible to derive quantum advantage in the near term [8].

We note that while alternate quantum computational approaches exist with relevance to transport, including quantum annealing [9-11], our focus here is on so-called "circuit-model" or "gate-based" implementations.

**Identifying transport optimization problems suitable for quantum approaches**

We have discussed the potential for quantum computing to enhance operational performance and customer outcomes in transport networks, but which problems are most suitable for obtaining enhancements in the near-term? We put forward several criteria for this assessment:

1. **Classical complexity:** The problem should be hard, requiring exponential-time to find the optimal solution using classical computing resources. Precisely, the problem should correspond to the optimization variant of an NP-complete decision problem. Furthermore the approximability of the problem should be low: heuristics should produce relatively low-quality solutions or have large runtime for large problem instances. Quantum optimization aims to provide high-quality solutions faster than classical approximate algorithms; this advantage is achievable sooner where the classical benchmarks are more limited.
2. **Practical impact:** The problem should be of sufficient impact and scale that a 1-2% classical suboptimality of conventional algorithmic approximations constitutes a substantial opportunity for value delivery. The most likely near-term improvements from quantum optimizations will be small enhancements in solution quality relative to the classical approach; ideally a problem should be chosen such that these improvements make a substantial difference in practice.
3. **Quantum-algorithm compatibility**: The mathematical representation of the problem and its constraints should be compatible with known approaches to optimization and (ideally) known quantum algorithms. A common form is the unifying problem of Quadratic Unconstrained Binary Optimization (QUBO), where the aim is to find the global minimum of a quadratic polynomial of binary variables. The QUBO framework provides a convenient embedding for NP optimization problems, and has a strong relationship to the quantum Ising model [12]. As such, QUBO formulations of optimization problems are directly compatible with an optimization algorithm known as QAOA, discussed further below. Although QUBO is the conventional approach for mapping optimization problems to quantum computers, it is not necessarily the optimal approach. Other approaches tailored to the specific optimization problem can in some cases reduce qubit requirements compared to the QUBO model.

4. **Resource efficiency**: A candidate problem should be able to deliver practically relevant results given practical near-term projections for available quantum resources (qubits and quantum gates). The required resources typically scale with the problem size or complexity, and this scaling will impact the expected crossover to outperforming classical solutions. Fewer required qubits and quantum gates lead to better algorithm implementation on error-susceptible quantum hardware.

**How to solve computational problems in transport with quantum technology**

Once an opportunity for quantum computing to impact critical transport network optimization problems is identified, the next step is to provide concrete pathways to quantum advantage - outperforming classical technology - in delivering useful transport solutions.

*Solution stack*
Figure 1 outlines the procedure for moving from a given transport problem to a high-quality solution, making expert use of a quantum computer. Integrating quantum technology into transport workflows requires expertise in a variety of domains, in addition to the hardware specialists who develop quantum computing systems. Quantum technology is undergoing active development at all layers of the stack. Progress in one layer should inform design at the other layers; a well-integrated stack ensures that larger-scale problems can be addressed sooner on quantum hardware, and that performance degradation through noise and error is minimized.

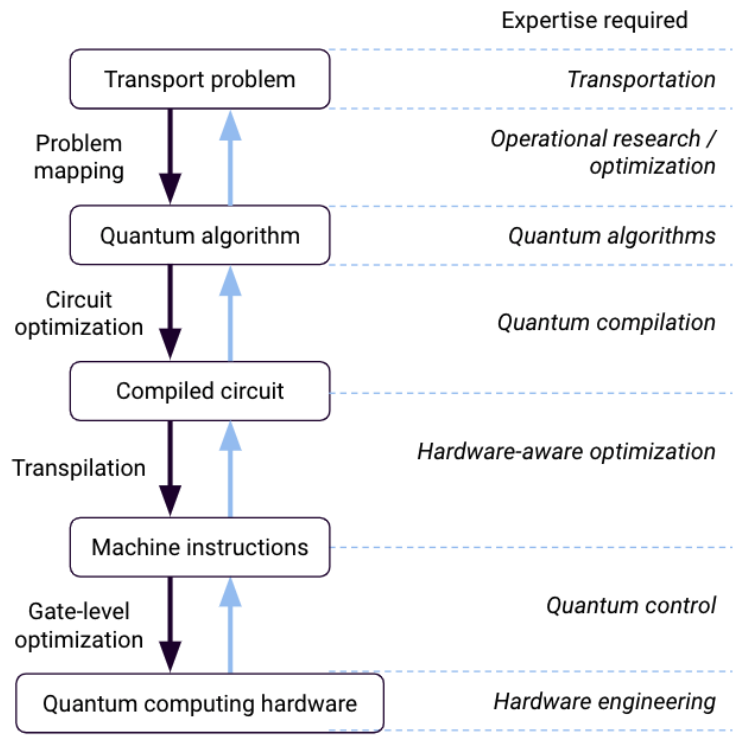

Figure 1: Quantum solution stack and associated expertise.

For quantum computing end-users entering at the top of the solution stack, engagement begins by identifying promising transport use cases through consultation between transport experts, operational research experts, and quantum algorithm designers. The chosen transport problem is then mathematically formalized (mapped) in a way that captures the key elements for the use case, and is compatible with resource-efficient implementation of a quantum algorithm.

The quantum algorithm, the next layer in the figure 1 solution stack, determines the method for solving the problem using quantum computing resources. Even with known algorithms there remains substantial opportunity for innovation in the design of modular building blocks used to implement the algorithms and deploy quantum circuits to represent the physical computation. Before execution, the quantum circuits must be optimized through a process known as compilation to simplify redundant quantum operations. The compiled circuit is then optimized (transpiled) to conform to the specific characteristics of the hardware (e.g. how different qubits may be physically connected). Hardware machine-specific instructions are passed through gate-level optimization to implement high-quality quantum logic on the hardware. In particular, it is possible to deploy "robust-control" strategies at this layer in order to build noise resilience in the constituent physical operation [13]. Appropriate consideration of the impact of noise and error in each step of this chain can deliver dramatically improved performance to the end user.

**Applying the methodology: quantum vehicle routing**

We have identified criteria for assessing problems that could obtain improved performance with quantum computing, as well as a workflow for delivering useful performance with near-term hardware. We now begin by analyzing the Capacitated Vehicle Routing Problem as a candidate for quantum augmentation, and introduce a novel problem encoding enhancing compatibility with near-term devices.

*Classical complexity*
In the CVRP, a fleet of vehicles with capacity constraints is deployed to pick up goods or passengers at a variety of locations. The problem is to determine the best set of routes for the vehicles to optimize an objective function such as travel time. CVRP is NP-hard, meaning no efficient classical algorithm for finding the globally optimal solution is likely to exist [9], and obtaining high-quality approximate solutions is a challenging problem.

Exact CVRP solutions can be found by sophisticated solvers for problems of size up to ~150 passenger locations with multiple vehicles, but execution times can become untenable depending on the problem specifications (for instance 11 days for a 284 node problem [14]). The best heuristic solution methods can reliably produce solutions within 1% of optimal/best known [15], and require on average 10 minutes of CPU time for problems with 50-199 passengers. While these algorithms - the product of many years of research - set a high bar, there is still room for improvement on these execution times, especially because problem sizes can dramatically exceed these size scales in urban networks.

*Practical utility and impact*
To address the second assessment criterion – the practical relevance of the CVRP – we first note that the CVRP can describe a variety of practical transport use-cases including first-mile/last-mile services, on-demand school services, emergency evacuation and freight. Qualitatively, minor improvements in solution quality can have substantial safety impacts for cases such as emergency evacuation. Quantitatively, the scale of existing use-cases ensures that even small gains in performance can amount to significant aggregated savings in cost, time, and environmental impact. In a real-world project involving the authors, a fast-moving consumer goods company reduced route distance by 10%, resulting in more than a million kilometers saved annually.

*Quantum-algorithm compatibility*
Existing literature has explored the potential of quantum optimization for the CVRP, such as in [9-11, 16]. The following workflow introduces our detailed problem assessment and enhancements across the solution stack.

We identify the Quantum Approximate Optimization Algorithm (QAOA), proposed by Farhi et al. [17], as relevant to the optimization problem underlying CVRP. The QAOA framework permits efficient representation of QUBO problems, with each binary optimization variable being mapped to the quantum information stored in a single qubit.

QAOA is broadly designed to provide high-quality approximate solutions for combinatorial optimization problems, with a number of its adaptations and implementations having the potential to provide near-term quantum advantage [18, 19] within a broad space of application areas. It can be considered as a generalization of the Grover quantum search framework [20, 21] oriented towards delivering quantum advantage by *efficiently* generating more desirable *approximate* solutions to combinatorial optimization problems, in a manner compatible with contemporary hardware constraints.

The QAOA operates in a hybrid algorithmic framework involving both classical and quantum computing components. The quantum circuit representing the problem employs an interleaved series of two modules: a problem or cost unitary, which applies a phase shift to solutions proportional to their quality, and a mixer unitary, which mixes quantum amplitudes between solutions to explore the search space. A classical optimization loop tunes these parameterized modules such that the combined exploration and solution-quality-dependent phase shifts make it increasingly likely to measure high-quality outputs. The quantum circuit is applied multiple times to obtain a number of measured states (candidate solutions) and construct an optimization cost associated with the variational parameters. The classical optimizer determines the next set of variational quantum circuit parameters until convergence or a stopping condition are reached. Finally, the best-obtained candidate solution is returned by the algorithm.

In the following we address the challenge of mapping the specific characteristics of the CVRP into QAOA in a resource-efficient manner.

*Resource-efficient problem mapping*
The mapping of an optimization problem to a linear programming or QUBO model has a direct impact on the required quantum computational resources. Given a mapping, the number of variables determines the number of qubits required (the circuit width), while the number of constraints will typically affect the number of sequential quantum gates applied in the execution of the algorithm (the circuit depth). The possible problem representations and the consequent impact on circuit structure should be taken into consideration for best results on near-term quantum hardware.

Here we introduce a novel 'binary' method for mapping the CVRP for quantum algorithms while reducing qubit resource requirements that (to our knowledge) has previously only been applied to the problem of clique partitioning [22].

The CVRP involves both assigning vehicles to particular passenger/goods locations, as well as ordering each vehicle's stops. We begin with the stop-ordering problem; for a given vehicle and list of stops, this can be described as the Traveling Salesperson Problem (TSP). Specifically, given a list of stops along with the relative distances/travel costs, the TSP asks for the shortest route that visits all stops exactly once, before returning to the depot.

We can think of potential solutions to the TSP as permutations of 'nodes' indicating the visit ordering. In [22], this problem is formulated with respect to variables indicating a particular node $i$ and its location $j$ in the visit cycle, such that the edge-based cost of traveling the route can be written as

$$c(x) = \sum_{u,v} W_{u,v} \sum_j x_{u,j} x_{v,j+1}$$

where $W_{u,v}$ is the edge-cost from node $u$ to node $v$, and the decision variable $x_{u,j}$ is 1 if node $u$ is visited $j$th, and 0 otherwise. Optimization of the route involves minimizing the above cost function, subject to constraints. The set of constraints ensure that each node appears exactly once in the visit cycle, and that each visit-cycle position is used exactly once. Expressing the decision variables as a matrix, the constraints are equivalent to requiring that the sum of each of the rows and columns must be equal to 1 – this is called a "one-hot" encoding.

The above QUBO-based formulation (provided in full in [22]) can then be cast to a quantum *problem Hamiltonian* (a mathematical representation of the quantum computational dynamics) via a one-to-one mapping from the binary variables to quantum operators acting on the corresponding qubits. For the $n$-node TSP, this mapping results in a problem Hamiltonian that is composed of $\sim n^3$ simple quantum operators, each of which acts on at most two qubits. The polynomial scaling of the Hamiltonian results in efficient quantum circuit construction and deployment even as the problem size grows.

For $n$ nodes (locations of passengers or goods), the one-hot encoding requires $n^2$ decision variables (and thus qubits) to index over nodes and visit-cycle positions. As an example of the type of design tradeoff discussed above, a more compact permutation encoding can be employed using $n \lceil log_2 n \rceil$ qubits, where the $k$th block of $\lceil log_2 n \rceil$ qubits encodes in binary the integer $i$, such that the $k$th node is in visit-cycle position $i$. Valid quantum states that represent a feasible tour are $|\sigma(1)\rangle |\sigma(2)\rangle ... |\sigma(n)\rangle$ for some permutation $\sigma \in S_n$. The binary encoding has the advantage of a quadratic reduction in qubit usage, at the expense of a problem Hamiltonian that introduces multi-qubit operations. These multi-qubit operations must then be decomposed into one- and two-qubit quantum gates for implementation on hardware, requiring careful optimization to avoid runtime overhead.

The CVRP can be thought of as an extension of the TSP to use multiple vehicles, with each vehicle having an associated capacity constraint that introduces an upper limit to the number of nodes it can visit. A variety of different methods are possible for extending the TSP to the CVRP in the linear programming/QUBO formulations. This is an area of ongoing research in both the classical and quantum optimization literature [10, 11, 23].

A simple CVRP formulation is to extend the one-hot TSP formulation above by adding an additional 'vehicle' index to each decision variable: $x_{u,j,v}$ is 1 if and only if node $u$ is visited at position $j$ by vehicle $v$. This leads to a cost function that is similar to the TSP, only differing by an additional sum over all vehicle indices. The constraints are related to the above TSP encoding – each node should appear in exactly one vehicle's visit-cycle, and the other constraints relating to capacity and visit order are formulated similarly. For $n$ total nodes assigned among $C$ (capacity) visit-cycle positions for $V$ vehicles, $n\,C\,V$ qubits are required.

We noted earlier that for the CVRP, finding the optimal solution starts to become unfeasibly slow or inaccurate from ~100 passenger nodes. For the 'Standard' CVRP mapping, encoding a problem constituting 7 vehicles and maximum vehicle capacity of 20 passengers requires ~14000 qubits. By contrast, applying the binary TSP encoding we introduce above to the full CVRP improves scaling to $C\,V\,\lceil log_2 n \rceil$, resulting in a requirement of only ~1000 qubits for the same problem instance. Since $n$ passengers are encoded in blocks of $\lceil log_2 n \rceil$ qubits for the binary mapping, a fixed number of available qubits can represent problems with additional passengers until $n$ passes the next power of 2. This provides an amortized exponential improvement in the ability to scale to large passenger numbers, which is illustrated in figure 2.

Although the required qubits are beyond currently accessible hardware systems, the recent growth rate of devices is exceptional. IBM has recently released their 127-qubit device, while their roadmap [24] identifies a device with over 400 qubits for later this year, and over 1000 qubits for 2023. These qubit numbers are shown in figure 2 to indicate the total numbers of passengers for CVRPs (with 7 vehicles and max. 20 passengers/vehicle) that fit on given hardware sizes. Our binary mapping reaches problem sizes where classical heuristics are suboptimal or impractically slow (the classical suboptimal CVRP threshold) for hardware devices on the IBM roadmap by ~2023, in stark contrast to the one-hot mapping. According to our criteria, the CVRP is a suitable candidate for near-term quantum advantage.

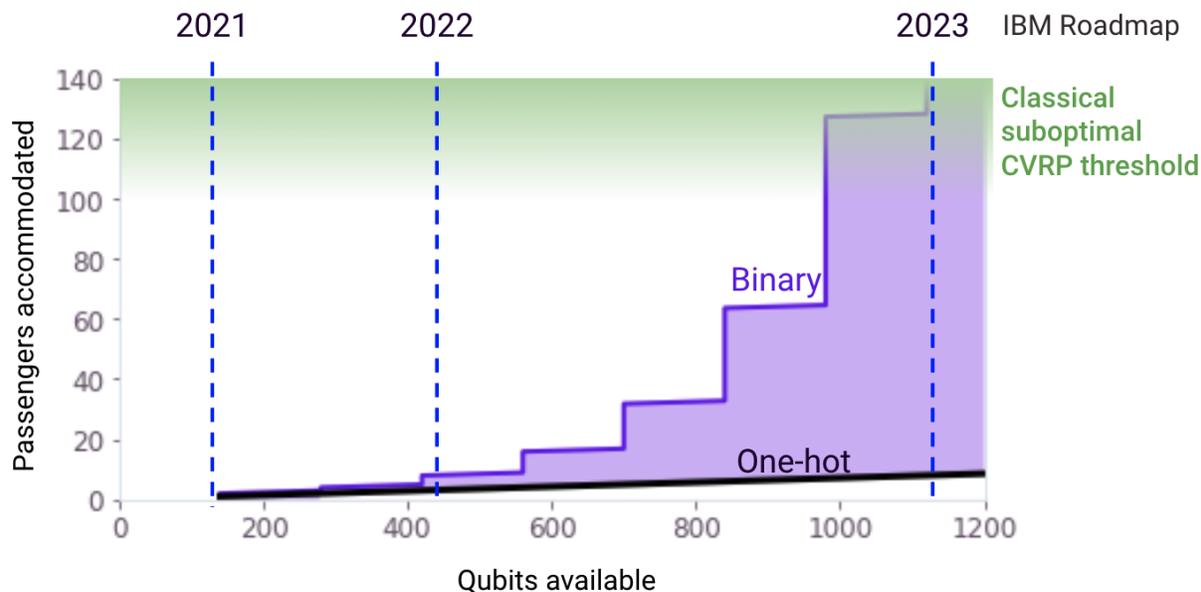

Figure 2: Standard one-hot mapping and Q-CTRL binary mapping resource requirements for CVRPs with 7 vehicles, max. 20 passengers/vehicle, and different numbers of total passengers.

**Efficient deployment on hardware**

Moving beyond general analyses of problem suitability and resource scaling, we now seek hardware-aware implementations which can deliver maximum performance from near-term, resource-constrained hardware. In this section we introduce two innovations that can enhance circuit success on real devices and provide explicit proof-of-concept demonstrations on small-scale quantum computers.

*Execution-efficient algorithm modules*
There are various adaptations of the QAOA algorithm [20, 25] as well as design decisions for its implementation. We focus here on quantum circuit design, which can be closely tied to the problem mapping and can lead to tradeoffs between different quantum resources which dramatically impact tractability.

In the mapping section above, we noted that valid solutions must satisfy certain constraints; in the QUBO standard formulation invalid solutions are discouraged by adding constraint-based penalty terms to the problem Hamiltonian. However, this is not the only method for implementing constraints. Instead, the invalid solutions can be eliminated from the search space entirely, implemented in QAOA circuits using "hard mixers" imposing hard constraints [25].

The conventional QAOA mixing circuit applies operations that modify the state of each qubit individually, with the variational parameter controlling the specific qubit manipulations. This mixer naturally explores the entire space of all binary strings. For the TSP, the number of binary strings is exponentially larger than the space of valid solutions, eliminating the chance of a quantum advantage when compared to a classical algorithm that only searches valid solutions. Instead, a hard mixer can be carefully designed to limit the algorithm's exploration to the 'valid subspace' of binary strings. Even for the more compact binary encoding, using Stirling's approximation we find that the feasible subspace is exponentially small (a factor of $\sim e^{-n}\sqrt{n}$ smaller for $n$ passengers) with respect to the dimension of the full state-space. This exponential reduction in search space leads to dramatic performance enhancement for the optimization algorithm when compared to the penalty-based approach [26], but generally brings with it the cost of expanded gate count (and hence pathways for error).

We have developed a new hard mixer implementation [27] which operates exponentially faster than previous implementations [20, 25] and is applicable to any QAOA problem involving permutations. The mixer operates by performing efficient quantum superpositions of swap operations between the $[log_2 n]$-sized registers that hold the indices of the permutation, where the amount of swapping is controlled by a variational parameter. This efficient hard mixer markedly improves upon the efficiency of QAOA for permutation-related optimization tasks such as the CVRP, and thus increases the potential for achieving near-term advantage.

*Execution performance enhancement in software*
In this example application, we employ Q-CTRL's Fire Opal software [28] to handle each element of the circuit deployment on real hardware.

The first element involves circuit optimization for effective hardware-aware implementation, as introduced in the solution stack discussion. Using Fire Opal's custom compilation procedure, a small-scale example circuit for routing a vehicle between a depot and two passenger nodes involves 4 total 2-qubit operations (quantum gates). The quantum circuit shown in figure 3 can be applied to determine the optimal sequence of two passengers for any input cost matrix, and thus can be applied as an optimal sub-problem solver for any vehicle routing problem that has a 2-passenger capacity limit.

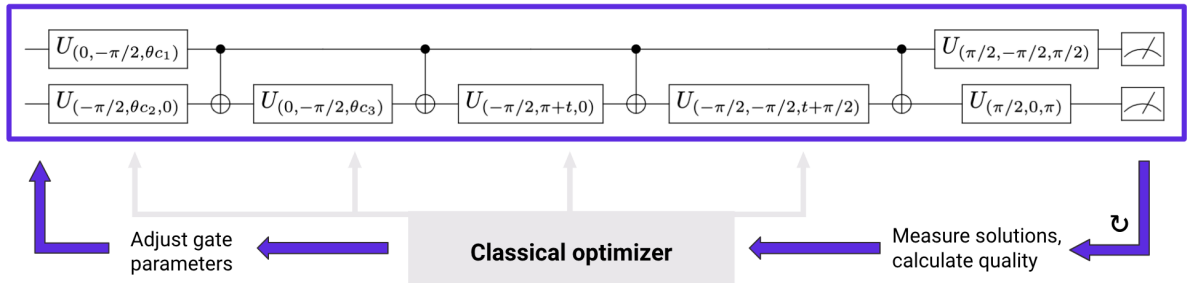

**Figure 3: An illustrative QAOA quantum circuit for solving any TSP instance involving a depot and two passenger nodes, using the binary encoding and a custom-designed hard mixer. Here $\theta$ and $t$ are variational circuit parameters, and the $c_i$ are instance-dependent parameters relating to travel costs.**

Particular hardware devices have their own specific error susceptibility to noise sources such as crosstalk, where operations intended to control one part of the device impact another part. After an optimized circuit is passed through the workflow, quantum firmware determines the optimal design of each individual quantum logic operation achievable on the device in order to return the target operation with minimal error susceptibility. Fire Opal automates gate design using closed-loop optimization routines in order to replace the most error-prone operations. The optimized circuits, composed of optimized gates, are then deployed to the device, and each measured output represents a potential solution to the CVRP. However the measurement procedure is also prone to error; readout-error correction is applied as a final step such that the measured circuit outputs are closer to the ideal error-free outputs.

In figure 4 the deployment results are shown for the 2-node problem represented by the quantum optimization process in figure 3. We plot the performance of the final quantum circuit, using the optimal values for the variational parameters as determined by the classical optimizer. The circuit performance corresponds to the probability of obtaining the correct routing result with respect to the ideal (error-free) circuit output. The Q-CTRL bars in the figure indicate the end-to-end solution design we have highlighted in this application example, including the binary mapping, problem-based algorithm design and firmware-enhanced deployment. This approach achieves 97% circuit performance in obtaining the correct routing result on IBM's 7-qubit *ibm_lagos* quantum computer, with error reduced by >20X over alternative implementations. In contrast, deployment of the compiled circuit with the standard one-hot mapping with penalties and the standard QAOA algorithm design (exploring the full state-space), denoted by the Standard (penalty) label, achieves 43% circuit performance. A constraint-based circuit identified in [25] using the one-hot mapping, denoted by Standard (hard mixer), involves higher circuit complexity and achieves 46% circuit performance. These results indicate that even for small problem sizes, appropriate problem implementation and execution can lead to appreciable increases in algorithmic problem success that can compound with problem size.

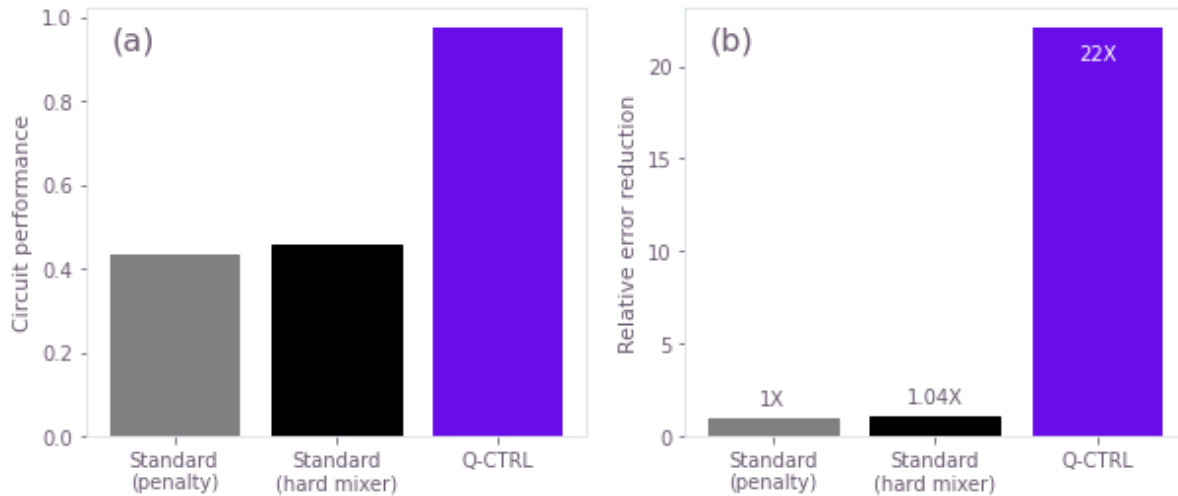

**Figure 4: QAOA circuit performance for different algorithm implementations, showing (a) the performance of the final quantum circuit after the optimization process shown in figure 3, and (b) the reduction in circuit error relative to the Standard (penalty) implementation.**

**Outlook for quantum technology in transport**

We conclude this article by summarizing our observations and results exploring the applicability of near-term quantum computing to transport optimization problems.

Unlike classical computing, where high performance can be obtained with relatively modular and independent infrastructure components, the existing quantum infrastructure has efficient algorithm design necessarily intertwined with hardware development. We expect that end-to-end solution design processes across the stack, supported by close interaction between different providers, will best advance quantum technology to achieve practical benefits in transportation. This claim is supported by our results of applying quantum optimization methodology to the Constrained Vehicle Routing Problem, where quantum optimization is applied to the problem of finding the optimal pickup order of a cluster of customers. Via careful selection of problem decomposition, mapping, compilation, and execution, we have demonstrated over 20X error reduction in the application example for an integrated solution stack over both a hard mixer circuit from the literature [25] and a smaller but non-scalable standard QAOA circuit.

The algorithm-module design displayed in this work for the CVRP involves superior scaling performance to alternatives in the literature, with quadratic qubit resource reduction, an exponentially-reduced search space compared to the standard QAOA implementation, and exponentially-reduced circuit depth scaling compared to existing hard mixers. These full-stack improvements bring quantum advantage substantially closer on near and medium-term hardware; devices on hardware roadmaps for 2023 offer sufficient qubits for obtaining quantum advantage. Further reduction of hardware errors and solution-stack development will be necessary to achieve this advantage in practice, and we demonstrated the use of error suppressing tools in execution workflows in our proof-of-concept experiments.

To extend the potential applicability of quantum advantage for near-term hardware, one approach to handle large-scale problems with intractable resource requirements is to decompose them into smaller, tractable subproblems. It is important to then assess the subproblems with respect to our criteria. For instance, resource reduction for the CVRP can be achieved by first applying a classical clustering step, which separates passengers into groups assigned to individual vehicles. The subsequent quantum optimization provides the optimal route for the vehicle among its assigned passengers [9] (TSP). In this case, however, although the TSP is an NP-hard problem (there are no efficient exact classical algorithms), the straight-line distance version of the problem admits a polynomial-time approximation scheme, and high-quality approximate solutions to the general problem can be obtained efficiently up to sizes of ~10,000 passengers [29]. As such, this decomposition method is not promising for near-term quantum advantage according to our criteria. Nonetheless, the TSP can be used as a useful benchmark for relevant algorithm performance that has natural extensibility to classically harder-to-approximate routing problems, while the available hardware scales towards the more challenging resource requirements of the CVRP. Alternative decompositions may identify sub-problems that are more classically challenging, bringing the threshold where quantum advantage becomes possible still nearer.

More broadly, transport operations involve a variety of services supported by data-driven tools; quantum technology should necessarily integrate with and extend existing classical solution architectures. Our analysis shows that even in the long term it is unlikely quantum solutions will fully supplant classical optimizations, but we have identified opportunities for quantum enhancement that can have material operational impact.

Such discussions highlight that flexible interfaces between classical and quantum computing stacks will therefore be critical to implementing quantum computing for operational and decision support systems. Potential areas for integration include classical data processing to obtain real-time, relevant and high-quality data, integration of quantum optimization as a routine within classical traffic simulation frameworks, and integrating classical AI/ML capabilities such as network forecasting to provide inputs for quantum optimization. We look forward to continued developments in quantum algorithms and their implementation on quantum hardware systems.

**Acknowledgements**

This work is supported by Transport for New South Wales as part of the proof of concept engagement: 'Quantum computing to impact transport-network optimisation problems'.

**References**
1. Barnhart, C. and Laporte, G. (2006) *Handbooks in Operations Research and Management Science: Transportation*. Elsevier.
2. Vogiatzis, C., & Pardalos, P. M. (2013). Combinatorial optimization in transportation and logistics networks. In *Handbook of combinatorial optimization* (pp. 673-722). Springer New York.
3. Shor, P. W. (1999). Polynomial-time algorithms for prime factorization and discrete logarithms on a quantum computer. *SIAM review*, 41(2), 303-332.
4. Grover, L. K. (1996, July). A fast quantum mechanical algorithm for database search. In *Proceedings of the twenty-eighth annual ACM symposium on Theory of computing* (pp. 212-219).

5. Durr, C., & Hoyer, P. (1996). A quantum algorithm for finding the minimum. *arXiv preprint quant-ph/9607014*.
6. Zhong, H. S., et al. (2020). Quantum computational advantage using photons. *Science*, 370(6523), 1460-1463.
7. Preskill, J. (2018). Quantum computing in the NISQ era and beyond. *Quantum,* 2, 79.
8. Bharti, K., et al. (2022). Noisy intermediate-scale quantum algorithms. *Reviews of Modern Physics,* 94(1), 015004.
9. Feld, S., et al. (2019). A hybrid solution method for the capacitated vehicle routing problem using a quantum annealer. *Frontiers in ICT,* 6, 13.
10. Irie, H., et al. (2019). Quantum annealing of vehicle routing problem with time, state and capacity. In *International Workshop on Quantum Technology and Optimization Problems* (pp. 145-156). Springer, Cham.
11. Borowski, M., et al. (2020). New hybrid quantum annealing algorithms for solving vehicle routing problem. In *International Conference on Computational Science* (pp. 546-561). Springer, Cham.
12. Glover, F., Kochenberger, G., & Du, Y. (2018). A tutorial on formulating and using QUBO models. *arXiv preprint arXiv:1811.11538*.
13. Carvalho, A. R. R., et al. (2021). Error-Robust Quantum Logic Optimization Using a Cloud Quantum Computer Interface. *Physical Review Applied*, 15, 064054.
14. Pessoa, A., Sadykov, R., Uchoa, E., & Vanderbeck, F. (2020). A generic exact solver for vehicle routing and related problems. *Mathematical Programming,* 183(1), 483-523.
15. Vidal, T., Crainic, T. G., Gendreau, M., & Prins, C. (2014). A unified solution framework for multi-attribute vehicle routing problems. *European Journal of Operational Research,* 234(3), 658-673.
16. Bennett, T., Matwiejew, E., Marsh, S., & Wang, J. B. (2021). Quantum walk-based vehicle routing optimisation. *Frontiers in Physics,* 9:730856.
17. Farhi, E., Goldstone, J., & Gutmann, S. (2014). A quantum approximate optimization algorithm. *arXiv preprint arXiv:1411.4028*.
18. Streif, M., et al. (2021). Beating classical heuristics for the binary paint shop problem with the quantum approximate optimization algorithm. *Physical Review A,* 104(1), 012403.
19. Zhou, L., et al. (2020). Quantum approximate optimization algorithm: Performance, mechanism, and implementation on near-term devices. *Physical Review X,* 10(2), 021067.
20. Bärtschi, A., & Eidenbenz, S. (2020). Grover mixers for QAOA: Shifting complexity from mixer design to state preparation. In *2020 IEEE International Conference on Quantum Computing and Engineering (QCE)* (pp. 72-82). IEEE.
21. Jiang, Z., Rieffel, E. G., & Wang, Z. (2017). Near-optimal quantum circuit for Grover's unstructured search using a transverse field. *Physical Review A,* 95(6), 062317.
22. Lucas, A. (2014). Ising formulations of many NP problems. *Frontiers in Physics*, 2(5).
23. Borcinova, Z. (2017). Two models of the capacitated vehicle routing problem. *Croatian Operational Research Review*, 463-469.
24. "IBM'S Roadmap For Scaling Quantum Technology". IBM, 2020, https://research.ibm.com/blog/ibm-quantum-roadmap.
25. Hadfield, S., et al. (2019). From the quantum approximate optimization algorithm to a quantum alternating operator ansatz. Algorithms, 12(2), 34.
26. Wang, Z., Rubin, N. C., Dominy, J. M., & Rieffel, E. G. (2020). XY mixers: Analytical and numerical results for the quantum alternating operator ansatz. *Physical Review A,* 101(1), 012320.
27. Q-CTRL Pty Ltd. (2021). *Quantum approximate optimisation*. Australian Patent Application No. 2021903901.
28. "Q-CTRL Boosts Quantum Algorithms By >25X In Benchmarking Experiments". Q-CTRL, 2022, https://q-ctrl.com/blog/q-ctrl-boosts-quantum-algorithms-by-greater-than-25x-in-benchmarking/.

29. Arnold, F., Gendreau, M., & Sörensen, K. (2019). Efficiently solving very large-scale routing problems. *Computers & Operations Research,* 107, 32-42.